\documentclass[fleqn]{pasj02}  
\usepackage[switch,mathlines]{lineno} 

\usepackage{mathptmx}
\usepackage{natbib} 
\usepackage{graphicx}

\jyear{2025}
\Received{2025/10/18}
\Accepted{2025/12/16}

\graphicspath{ {./figures_ando/} } 

\begin{document} 

\title{ Non-adiabatic Effect on Convective Mode }

\author{
 Hiroyasu \textsc{Ando},\altaffilmark{1,2} \email{ando.hr@iris.ocn.ne.jp} 
}
\altaffiltext{1}{National Astronomical Observatory of Japan, National Institutes of Natural Sciences, 2-21-1,Osawa, Mitaka, Tokyo, 181-8588, Japan}
\altaffiltext{2}{Department of Astronomy, Graduate University for Advanced Studies, Shonan Village, Hayama, Kanagawa, 240-0193, Japan}



\KeyWords{convective mode -- radiative thermal conduction -- oscillatory convection -- wave energy -- solar model}  

\maketitle

\begin{abstract}
 The systematic analysis of non-adiabatic effect on convective mode has been conducted using wave energy 
relation. In the adiabatic analysis, the "propagation diagram" for convective mode is proposed as a
useful tool to see its behavior.  In the non-adiabatic analysis, 
 it is found that for strongly non-adiabatic case, a monotonically growing convective mode becomes 
 oscillatory.   
  In this phase, the radial displacement and the distribution of wave 
  energy show only one bump, in which the distribution of entropy energy $e_S$ almost overlaps  with
   the distribution of gravity energy $e_g$.  Entropy energy $e_S$ seems to act as  potential energy of oscillatory convection.   In addition to this, this change occurs not gradually, but abruptly with change of non-adiabatic
   indicator.
\end{abstract}


\section{Introduction}
The oscillatory convection  was discovered by \cite{shiba1981}  in the global analysis for highly luminous ($L=10^5 L_{\Sol}$) stars.  In this study, a new type of non-adiabatic unstable modes (called as "A-mode") was 
 found, in which the growth rate (imaginary part of the complex eigenvalue)  is comparable with the frequency 
 (real part).
 So these modes could not be classified as p-, f-, and g-modes. To make clear what kind of mode it is, they
  performed the numerical experiment for 
"A-mode", and examined how it changes its behavior by changing non-adiabaticity indicator $C_2$  in the basic linear equations of non-adiabatic nonradial oscillations \citep{ando1975}.  Eventually,  "A-mode" was revealed to be a 
convective mode (i.e., g$^{-}$-mode) in the adiabatic limit (large  $C_2$).  They concluded that "A-mode" is an oscillatory convection originated from convective mode.  However, they stopped the further systematic study of why
and how a convective mode becomes oscillatory due to strong non-adiabaticity.  After their work, no systematic study of non-adiabatic effect on convective mode has been conducted, as far as we know.  

Three decades later, oscillatory convection was paid an attention to from observational point of view.
 \citet{saio2011} and \citet{saio2015} have proposed oscillatory convection for the explanation of variability
  in the luminous  variable stars.  \citet{latter2016} have also conducted numerical calculations of overstable convection as a dynamical movement in  protoplanetary discs.  

So far,  in the local analysis,  \citet{cowling1957} and \citet{kato1966} studied the overstability of convective
mode under the super-adiabatic medium stabilized by the magnetic fields and by the chemical composition 
gradient, respectively.   But strictly speaking, this kind of instability is different from findings by \citet{shiba1981},
since  the super-adiabatic  medium is not stabilized in their situation. The systematic study
following their work is, therefore, highly expected.

This paper is devoted  to the essential problem of why and how a convective mode becomes oscillatory 
with use of wave energy (Eq. 25.12 given in \citet{unno1989}).  The present solar model is used as an
equilibrium model throughout this work. 
 In section 2, the basic aspect of adiabatic convective mode is reviewed following the global analysis done by \citet{hart1973}.  Then,  "propagation diagram" for
 g$^{-}$-mode similar to that for p-, f-, and g-mode is newly introduced to discuss their features.  In 
section 3,  the non-adiabatic analysis of convective mode is developed to understand how a monotonically
growing convective mode becomes oscillatory  under the thermal conduction through radiation.   
Section 4 is summary and discussion.

\section{Adiabatic analysis}\label{sec:2}
\indent As for adiabatic convective mode we use  the basic equations for the adiabatic nonradial
oscillations given in the Chapter 18 (Eq. $18.14 \sim 18.17$) of \citet{unno1989}. 
The solar model at the present time
is calculated as an equilibrium model by \citet{pacz1969} program, of which parameters are
 $T_e = 5770K,  R_{\Sol}=6.96 \times 10^{10} $ cm , the ratio of the mixing length to the pressure scale 
 height $l/H_p= 2.0$ , chemical composition $X= 0.7$, $Z= 0.03$, and the envelope mass $0.1 M_{\Sol}$ .
\subsection{Propagation diagram}\label{ssec:21}
\begin{figure}
	\includegraphics[width=\columnwidth]{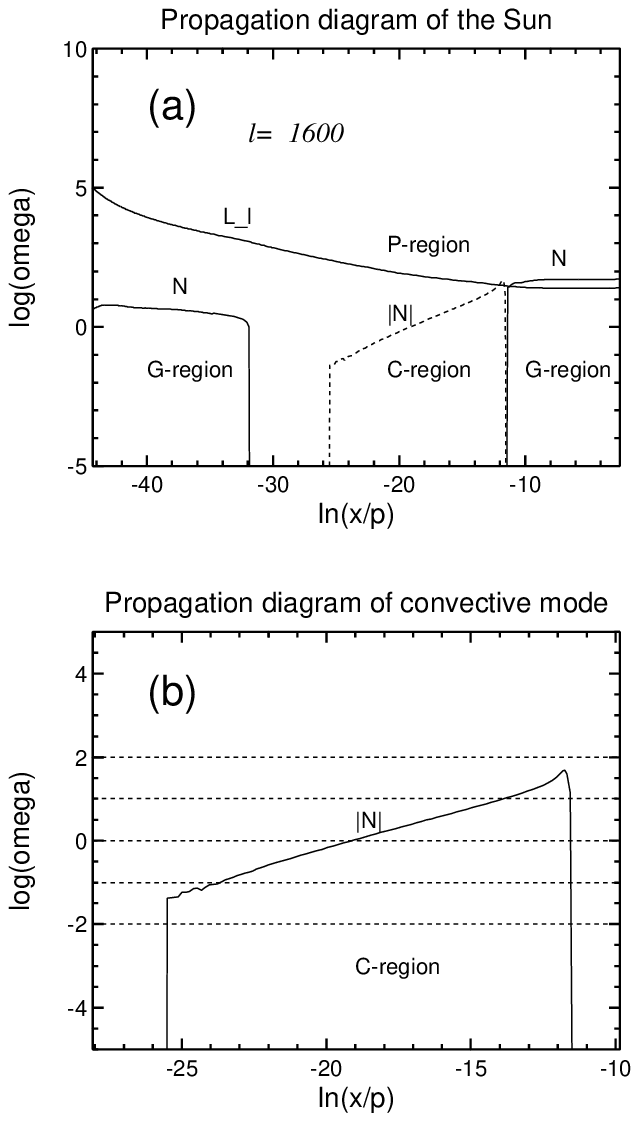}
 \caption{(a):The propagation diagram of the Sun. Abscissa is ln(x/p), where
    x is fractional radius of the Sun ($r/R_{\Sol}$) and p is pressure. 
   The left end is the center of the solar core and the right end is the solar surface.
     Ordinate is non-dimensional frequency
      ($\omega = \sigma / \sqrt{GM_{\Sol}/R^3_{\Sol}}$).
    Solid lines designated by $L_l$ and $N$ indicate Lamb frequency and 
    Brunt-$\rm{V}\ddot{a}is\ddot{a}l\ddot{a}$ frequency, respectively.
      p-mode and g-mode can propagate in the P-region and  in the G-region,
     respectively.
    $|N|$ is given by dashed line.  Convective mode (g$^-$-mode) is confined in the 
    C-region.  
    \newline  
    (b): Enlarged propagation diagram of convective mode. A group of horizontal broken lines
    is a good indicator to see a confined region for a given $\omega_I$ of  g$^-$-mode. \\
    {Alt text: Two line graphs. In the upper panel, x axis shows radial position of the Sun. 
     y axis shows non-dimensional frequency measured with free fall timescale of the Sun. 
     The upper panel describes in what region a mode is trapped.  In the lower panel,
      C-region is enlarged}
     }
    \label{fig:propagation_figure}
\end{figure}

We understand that propagation diagram is a very useful tool to see their trapping region for p-, f-, and g-mode,
 that is, we can judge how a particular mode of nonradial oscillations behaves in the interior of a stellar model.
  This diagram is derived from the local dispersion relation of  which derivation should consult the paper \citep{osaki1975}  if necessary.   Here   
 the temporal and the spatial variation of perturbations are assumed to be exp($i\sigma t$) and exp($ik_r r$), respectively.   The resulting dispersion relation is  given by
\begin{equation}
k^2_r = \frac{(\sigma^2 - L^2_l)(\sigma^2 - N^2)}{\sigma^2 c^2} ,
 \label{eq:dispersion}
\end{equation}
 where the Lamb frequency and the Brunt-$\rm{V}\ddot{a}is\ddot{a}l\ddot{a}$ frequency
 are given, respectively in the following,
\begin{equation}
 L_l^2 = \frac{\ell(\ell+1)c^2}{r^2} , 
 \label{eq:Lamb}
 \end{equation}
 and
 \begin{equation}
 N^2= g \left( \frac{1}{\varGamma_1}\frac{dlnP}{dr} - \frac{dln\rho}{dr}\right) ,
 \label{eq:Brant}
 \end{equation}
 where $c$ is sound speed ,  
 The propagation diagram of the Sun is given in Fig.~\ref{fig:propagation_figure}, where characteristic frequencies
 $L_l$ and $N$ are depicted as a function of radial coordinate $ln(x/p)$.  In Fig.~\ref{fig:propagation_figure}(a), the solar inner
 core  ranges from ln(x/p)= -43.3 to  ln(x/p)= -34.1, and its envelope
  from ln(x/p)= -34.1 to ln(x/p)= -2.5, where x is fractional radius of the sun ($r/R_{\Sol}$)
  and p is pressure.

For a particular mode (given $\sigma$), a corresponding horizontal line in Fig.~\ref{fig:propagation_figure}(a)
 intersects $L_l$ and/or $N$ curves.  If $\sigma^2 > L^2_l, N^2$ or $\sigma^2 < L^2_l, N^2$ in 
Eq.~(\ref{eq:dispersion}), we obtain $k^2_r > 0$ 
and the solutions of the wave equations in this case present propagating waves. So the propagation diagram
(Fig.~\ref{fig:propagation_figure}(a)) shows  in which portion a particular wave can propagates. 
In the portion with $\sigma^2 > L^2_l, N^2$, a wave has p-mode nature. In case of $\sigma^2 < L^2_l, N^2$,
a wave has g-mode character.  On the other hand,
   if $N^2 > \sigma^2 > L^2_l$ or, $L^2_l > \sigma^2 > N^2$, we get $k^2_r < 0$.  The waves are then
   non-propagating as a running wave and they are called evanescent waves.

 In the case of convective mode, $\sigma $ is pure imaginary, and so $\sigma$ is 
 replaced by ($ i \sigma_I$) in Eq.~(\ref{eq:dispersion}), which is reduced to
 \begin{equation}
k^2_r = - \frac{(\sigma^2_I + L^2_l)(\sigma^2_I + N^2)}{\sigma^2_I c^2} ,
 \label{eq:dispersion_convection}
\end{equation}
In order for $k_r $ to be real,  it should be $\sigma^2_I < -N^2$. This condition is  satisfied only in the convective region ($N^2 < 0$). Here we  define $ |N|= \sqrt{-N^2}$ and 
plot it  by dashed line in Fig.\ref{fig:propagation_figure}(a).  As the growth rate  $\sigma_I$ should be
 below $|N|$ line, a convective mode is confined in the C-region. Thereafter, we  call it  as "propagation diagram" for convective mode.  Enlarged propagation diagram is shown in Fig.\ref{fig:propagation_figure}(b) to see the
 trapped feature in more detail.

\subsection{Wave energy}\label{sec:wave_energy}
Wave energy is a crucial tool to understand the nature  not only of conventional mode, but also 
of convective mode. 
The wave energy ($e_w$)
per unit mass is given  by Eq.(25.12) in \citet{unno1989}  such as,
 \begin{multline}
e_w = \frac{1}{2} \Biggl\{ \mathbf{v}^2 +  \left( \frac{p'}{\varrho c}  \right)^2  \\
+ \left( \frac{g}{N} \right)^2 \left[ \left( \frac{P'}{\Gamma_1 p}
 - \frac{\varrho'}{\varrho} \right)^2     
- \frac{\nabla}{\nabla_{ad}} \left(  \mathit{v}_T \frac{\delta S}{C_p}  \right)^2  \right] \Biggr\},
 \label{eq:energy_total}   
 \end{multline}   
where $\nabla$ is temperature gradient ($dlnT/dlnP$) and $\nabla_{ad}$ is adiabatic temperature
gradient ($\partial lnT/\partial lnP)_S$, and $\mathit{v}_T =  ( \partial  ln \varrho/\partial lnT)_p $ and the other
symbols have the usual meanings.   The variables with prime and $\delta$ represent Eulerian and
Lagrangian perturbations, respectively.  

To proceed further analysis, we transform each term in $e_w$ into a more convenient
form.  Using the displacement vector
 $\mathbf{\xi}= (\xi_r, \xi_h \partial/\partial \theta, \xi_h \partial / \mathrm{sin}\theta \partial\phi ) Y^m_l(\theta, \phi)e^{i\sigma t} $ and the relation $\mathbf v =  i \sigma \mathbf{\xi}$ ,  
 each energy term is reduced  after some manipulations in what follows,
 \begin{equation}
 e_k  =   \frac{1}{2}  \mathbf{v}^2 = \frac{1}{2} \Bigl\{  \left( i\sigma \xi_r \right)^2 +\ell(\ell+1)  
 \left( i\sigma \xi_h \right)^2
  \Bigr\},  
 \label{eq:energy_kinetic}
 \end{equation}
\begin{equation}
 e_p  =   \frac{1}{2}  \left( \frac{p'}{\varrho c}  \right)^2, 
 \label{eq:energy_pressure}
 \end{equation}
\begin{equation}
e_g   =   \frac{1}{2}  N^2 \Bigl( \xi_r - \mathit{v}_T \frac{g}{N^2} 
\frac{\delta S}{C_p} 
  \Bigr)^2, 
 \label{eq:energy_gravity}
 \end{equation}
\begin{equation}
e_S  =   - \frac{1}{2} N^2 \frac{\nabla} {\nabla_{ad}} \Bigl(  
\mathit{v}_T \frac{g}{N^2} 
\frac{\delta S}{C_p}
  \Bigr)^2 .
 \label{eq:energy_entropy}
 \end{equation}
 Here  the kinetic energy $e_k$ is divided into radial component $e_{kr}$ (the first term of 
 Eq.~(\ref{eq:energy_kinetic})) and horizontal one $e_{kh}$ (the second term), and then
 $e_k = e_{kr} + e_{kh}$.  Here we call $e_p$, $e_g$, and $e_S$ as acoustic energy, gravity energy, and
 entropy energy, respectively.  We also define the integrated energy $E_k$, $E_{kr}$
 $E_{kh}$, $E_p$, $E_g$, and $E_S$ over the relevant
 region for the corresponding  energies per unit mass, (i.e., 
 $E_k= \int e_k dM_r$,...,etc.).

 $e_k$ is the kinetic energy of wave (or motion) and the other terms represent the potential energies corresponding to the restoring forces. When $e_p$ is dominant (i.e., p-mode ),  
 its wave energy is 
 equi-partitioned into  kinetic energy $e_k$ and acoustic  energy $e_p$ .
 Provided that $e_g$ is dominant in the adiabatic case (i.e., $\delta S = 0$),
the internal gravity mode ($N^2 >0$: g-mode) or the convective mode 
 ($N^2 < 0$: g$^-$-mode) appears.    In the former case, the gravity energy $e_g$ 
is potential energy for the internal gravity wave. In the latter case, the  energy $e_g$
becomes the source term for convective mode ($g^-$-mode). 

 In case of  $\delta S \neq 0$, the entropy energy $e_S$ appears. 
 When the term in the square bracket in $e_S$ is small enough comparing with the first term in 
 the square bracket of $e_g$,  the argument for $g$-mode or $g^-$-mode is not so different from that of
 the adiabatic case mentioned above.  The effect of entropy energy $e_S$ is comparatively small.
 
 When the second term in the square bracket in $e_g$ dominates over the radial displacement ($\xi_r$),
 the relation $e_S \approx -\nabla/\nabla_{ad}e_g$ is realized. As a factor $\nabla/\nabla_{ad} \sim 1$ in
 order of magnitude in the relevant region,
  it may suggest that they have  alternate roles, that is, when one plays a role
 of source term, the other plays a role of potential energy, and vice versa. The discussion for 
 $g$-mode or $g^-$-mode is absolutely different from that of the adiabatic case.  
 We will discuss the meaning of these terms in more detail in non-adiabatic analysis.
\subsection{Convective mode  }
\begin{figure}
	\includegraphics[width=\columnwidth]{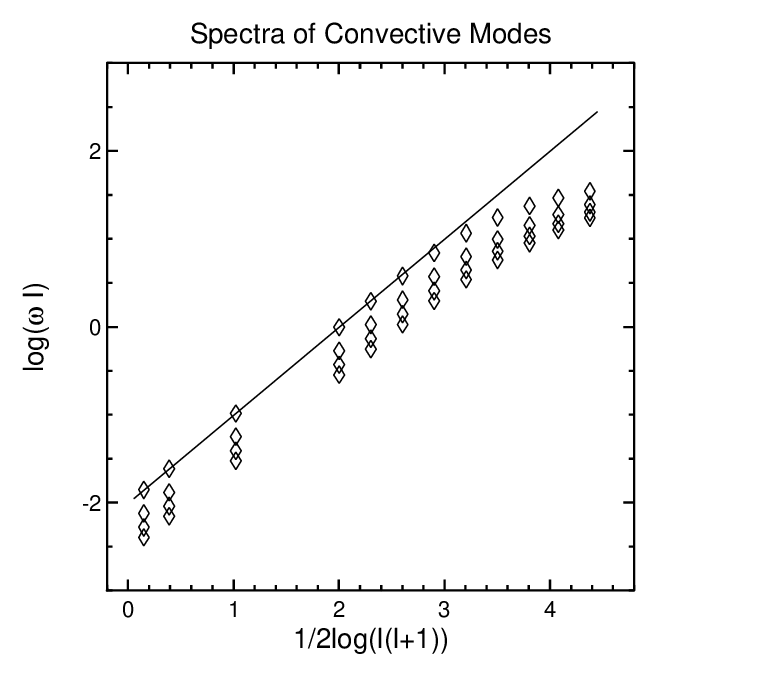}
    \caption{The relation of growth rate $\omega_{I}$ with spherical harmonics index $\ell$ is shown.  
    The growth rates of g$^{-}_0$-, g$^{-}_1$-,
    g$^{-}_2$-, and g$^{-}_3$-mode are plotted for each $\ell$ from top to bottom, respectively .  
    One straight line with slope 1 runs through the growth rate of g$^{-}_0$-mode for $\ell$=1.  \\
    {Alt text: x axis shows logarithm of spherical harmonics indicator el .
     y axis shows logarithm of growth rate of a mode. }
      }
    \label{fig:frequency_figure}
\end{figure}

Judging from the position of C-region in Fig.\ref{fig:propagation_figure}(a), we set the 
integration region to  [-28.0, -2.5] in ln(x/p).  Here the convective modes (the fundamental mode g$^-_0$ and three overtones g$^-_1$, g$^-_2$, and g$^-_3$) are calculated  for each $\ell$, (i.e.
$\ell=1, 2, 10, 100, 200, 400, 800,1600,3200, 6400, 12000$, and $24000$).  The results are summarized 
in Fig.~\ref{fig:frequency_figure}, in which non-dimensional growth
 rate $\omega_I = \sigma_I /\sqrt{GM_{\Sol}/R^3_{\Sol}}$ is plotted.
 
In this figure, one straight line with slope 1 running through a point of g$^-_0$-mode 
  for  $\ell=1$ is drawn.  The points for g$^-_0$-mode fit well in the lower $\ell$ region to this line.   It is easily derived from Eq.~(\ref{eq:dispersion_convection}) that $\sigma^2_I$ is proportional to
$\ell(\ell+1)$ in case of  $\sigma^2_I  <<  L^2_l, |N|^2 $, which was already pointed out 
by \citet{hart1973} .  It can be also applicable to growth rates of other overtones. 
But they gradually leaves this line towards larger $\ell$ and is saturated because the condition
of $\sigma_I < |N|$ should be satisfied. 

\begin{figure*}
	\includegraphics[width=1.0\columnwidth,angle=90]{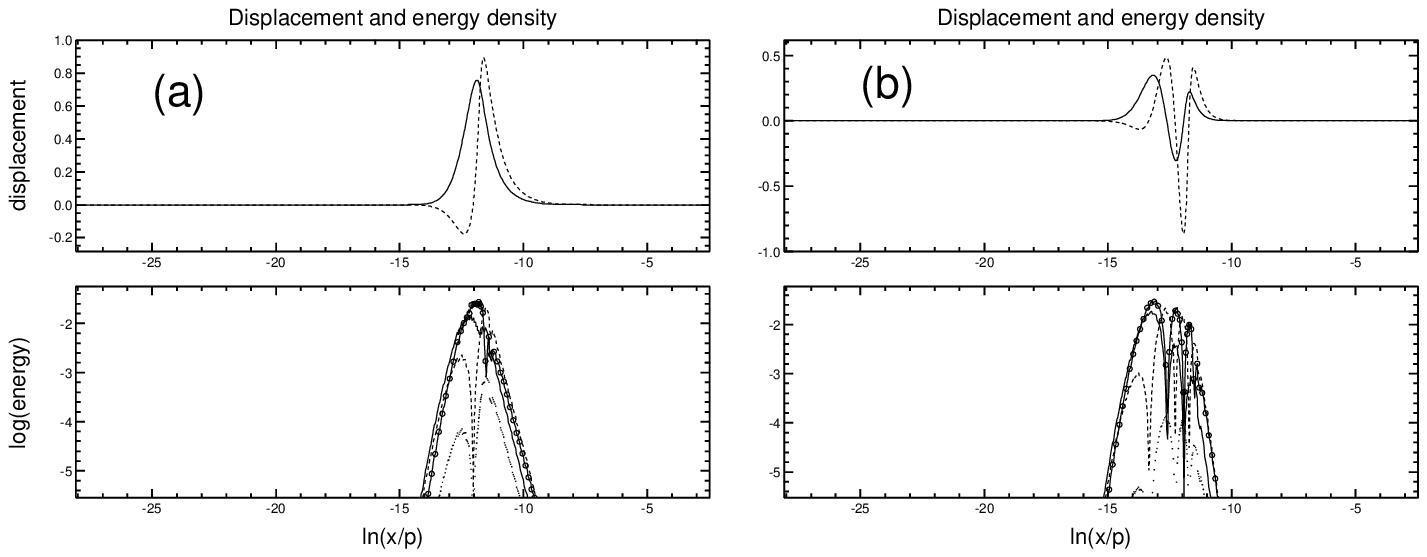}
    \caption{(a): Upper panel shows displacement $\xi_r/r$ (solid line) and 
    $\sqrt{l(l+1)}\xi_h/r$ (broken line) for $g^-_0$-mode  with $\ell= 6400$. In lower panel,   
    radial distribution of energies $e_{kr}$(solid line), $e_{kh}$(broken line), 
    $e_p$(dotted line), and $e_g$ (solid line with $\bigcirc$).   The abscissa is ln(x/p) and
    the ordinate is arbitrary. 
    \newline (b): The same plot as in figure (a) for $g^-_2$-mode  with $\ell=6400$.     \\
    {Alt text: The left panel shows how radial displacement and potential energies of 
    fundamental convective mode behaves in the radial direction of the Sun.  
    x axis shows radial position of the Sun.  y axis in the upper panel 
    shows arbitrary unit. In the lower panel, logarithm of energy. 
    The right panel shows the same line graph for the second overtone mode. }
     }
    \label{fig:cnv_adia_figure}
\end{figure*}

The the radial displacement  ($\xi_r/r$) (solid line) and the horizontal one ($\sqrt{\ell(\ell+1)}\xi_h/r$) 
(broken line) for g$^-_0$-mode and g$^-_2$-mode with $\ell =6400$ in the upper panels of Fig.~\ref {fig:cnv_adia_figure}, respectively. 
The radial distributions of energies 
 ($e_{kr}$, $e_{kh}$, $e_p$, and $e_g$)  are also shown  in the lower panels of 
 Fig.~\ref {fig:cnv_adia_figure}.   It should be noted here that the absolute values of energies 
 are plotted, regardless of their signs. 
 
 We can see from "propagation diagram" of convective mode in 
 Fig.~\ref{fig:propagation_figure}(b) that the radial and the horizontal displacements for g$^-_0$-mode and
  g$^-_2$-mode  are well confined between intersected points in C-region by horizontal lines of 
   their growth rates $\omega_I$, respectively.

From the upper panel  of Fig.~\ref{fig:cnv_adia_figure}(a) (g$^-_0$-mode), convective motion
in a box-like cell cut off by top and bottom boundaries, and vertical boundaries specified by
 indices ($\ell , m$) of the spherical harmonics $Y^m_l(\theta, \phi)$ is easily imagined.   A  peak of radial displacement (solid line) corresponds to zero
point of the horizontal displacement (broken line).  At both feet of  the  radial displacement,
the local minimum  and maximum of horizontal displacement appear.  These features suggest circular
motion of a gas element between the present cell and a neighboring cell. 
In addition to this, a peak of $e_{kr}$ of g$^-_0$-mode corresponds to the local minimum of
 $e_{kh}$ as shown in the lower panel of Fig.~\ref{fig:cnv_adia_figure}(a). This fact also
 supports this interpretation. 

In Table~\ref{tab:cnv_adia_table}, growth rate $\omega_I$, radial component $E_{kr}$ and horizontal component $E_{kh}$ of the total kinetic energy $E_k$, the total acoustic potential energy $E_p$, and the total
gravity energy $E_g$ are given for four modes, where  
 all energies are normalized by $E_k$.
Here the magnitude of $E_k$ and $E_g$
 is almost same but with opposite sign for all modes.  This comes from the fact that time derivative of
  wave energy $e_w$ 
 is zero owing to the adiabaticity.  Then,
the gravity energy $E_g$ is transferred into the kinetic energy $E_k$
and convective motion monotonically grows. 

One important point is the distribution of gravity energy $e_g$  shown by solid line with open circle
 in the lower panels of
Fig.~\ref{fig:cnv_adia_figure}.   ~~$g^-_0$- and $g^-_2$-mode have one bump and three bumps in
their distributions, respectively,  which corresponds to number of nodes for  each  eigenfunction
shown in the upper panels of  Fig.~\ref{fig:cnv_adia_figure}.  The radial displacement $\xi_r$ is alone
 included  in the square bracket of the gravity energy $e_g$.

\begin{table}
	\centering
	\caption{Characteristics of g$^-_0$-, g$^-_1$-, g$^-_2$-,  and g$^-_3$-mode
	 with $\ell=6400$ are given. $E_{kr}$ and $E_{kh}$ are radial  and horizontal 
	component of kinetic energy $E_k$, respectively.  All energies 
	are normalized by $E_k$.
	}
	\label{tab:cnv_adia_table}
	\begin{tabular}{lccccr} 
		\hline
		Mode & $\omega_I$ & $E_{kr}$ & $E_{kh}$ & $E_p$ & $E_g$ \\
		\hline
		 g$^-_0$ & -23.7 & 0.618 & 0.382 & 0.015 & -1.020 \\
		 g$^-_1$ & -14.2 & 0.518 & 0.482 & 0.006 & -1.006 \\
		 g$^-_2$ & -10.8 & 0.459 & 0.541 & 0.004 & -1.004 \\
		 g$^-_3$ & -8.93 & 0.417 & 0.583 & 0.003 & -1.003 \\
		\hline
	\end{tabular}
\end{table}
%
\section{Non-adiabatic analysis}
The basic equations for non-adiabatic nonradial oscillations given by \citet{ando1975}
are used, in which Lagrangian perturbation of the divergence of convective flux is neglected
 (i.e., $\delta(\nabla \cdot \mathbf{F_c})=0$).  The derivation of these basic equations is described
 in detail in the Appendix 1 of \citet{ando1975}.
 The same  boundary conditions are employed except for 
 the outer mechanical boundary condition.
 
\begin{figure*}
       \includegraphics[width=1.1\columnwidth,angle=90]{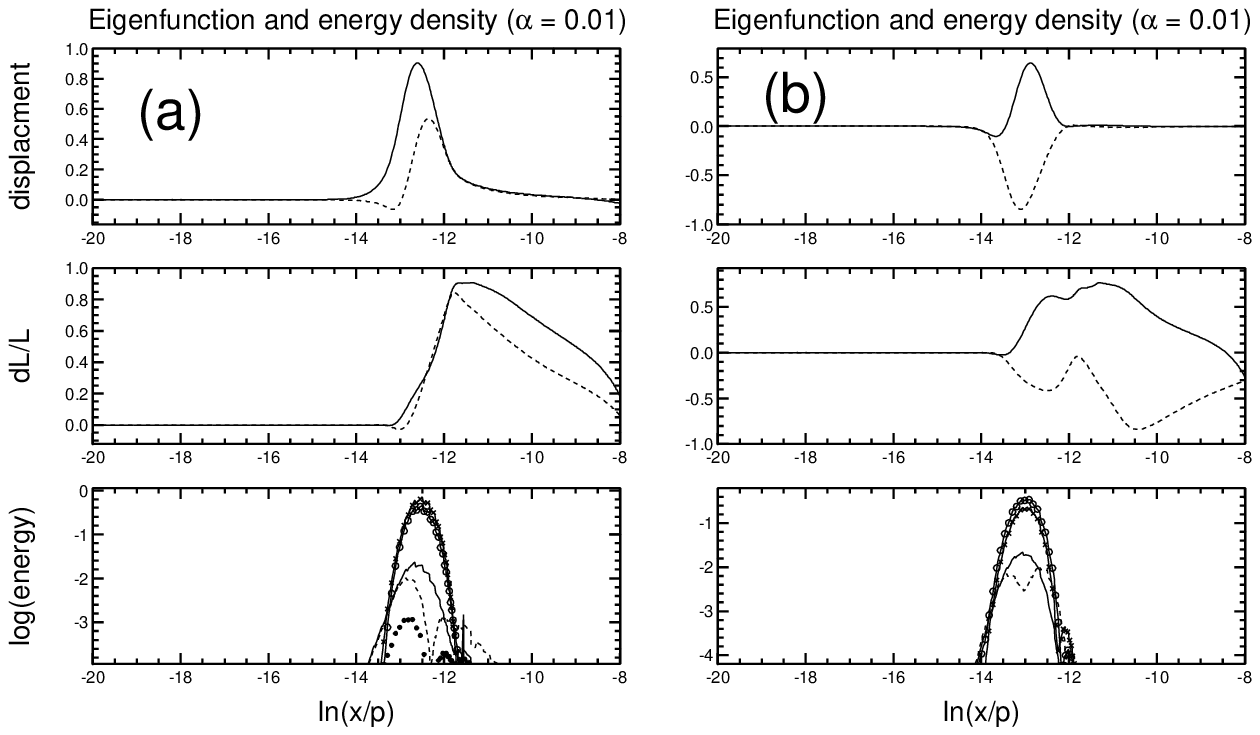} 
    \caption{(a): Non-adiabatic  displacement and  energy density of $g^-_0$-mode 
    with $\ell= 6400$ for $\alpha= 0.01$.   
    Solid  and broken lines in the upper panel indicate the real part and the imaginary part of radial displacement, 
    respectively.  Middle panel shows the luminosity perturbation ($dL/L$). In the lower panel, a line with 
    mark (x)  presents the distribution of entropy energy $e_S$.  
    \newline   
    (b): Non-adiabatic  displacement, luminosity perturbation, and  energy density distribution for $g^-_2$-mode. \\
    {Alt text: Two line graphs. These graphs show the behaviors of oscillatory convection.}
     }
    \label{fig:cnv_nadi_c-2_figure}
\end{figure*}

As shown in the propagation diagram (Fig.~\ref{fig:propagation_figure}), an eigenfunction with
 $\omega_R \neq 0$ and higher $\ell$ can propagate as a $g$-mode in the outer G-region and run 
 away from the surface. 
 So the zero boundary condition $\delta P=0$ is set as a mechanical boundary condition.   The diffusion approximation of radiation transfer is also adopted as the thermal conduction.
 To compare with the adiabatic analysis, the case of $\ell = 6400$ is calculated here. 
 Integration region is limited to ln$(x/p)$= [-28.0, -8.0]. The inner and the outer  boundaries are selected to avoid mixed mode with g-mode in  the G-regions (see Fig.~\ref{fig:propagation_figure}(a)),
 because we concentrate on examination of properties of convective mode.  As shown later in 
 Figs.~\ref{fig:cnv_nadi_c1_figure} and \ref{fig:cnv_nadi_c-2_figure}, an eigenfunction of convective mode is,
 in fact, well trapped in C-region. 

  \citet{shiba1981} showed  that a mode with unusually large growth rate ($|\omega_I| \sim \omega_R$) tends to a convective mode in the  adiabatic limit. They replace  the coefficient $C_2$
  in the basic equations by $\alpha C_2$.  Here $C_2$ is related with the ratio of local thermal time scale to dynamical time scale such as $\tau_{th}/\tau_{dyn}= | \omega |C_2/V$ [i.e., equation (13) in \citet{shiba1981} ],
  where $V$ is homologous invariant ($-dlnP/dlnr$). 
    By changing a parameter $\alpha$, they artificially
  change the degree of non-adiabaticity.  For instance, $\alpha=0$ corresponds to the extreme
  non-adiabaticity (i.e., isothermal perturbation).  In the limit of $\alpha \to \infty$, a perturbation
   behaves almost adiabatically.
  Now we calculate convective modes for three cases (
  $\alpha = 1.0, ~ 0.1, ~ 0.01 $).  The results are summarized in Table~\ref{tab:cnv_nonadi_table}. 
 

\begin{table*}
	\centering
	\caption{Characteristics of non-adiabatic convective modes, g$^-_0$-, g$^-_1$-, 
	g$^-_2$-,  
	and g$^-_3$-mode with $\ell=6400$ are given for each non-adiabatic indicator 
	$\alpha= 1.0, ~ 0.1, ~ 0.01$.  The parameter  $(\alpha\tau_{th}/\tau_{dyn})_{pk}$
	is a value at the peak of the distribution of $e_S$.
	 ~$E_{kr}$  and $E_{kh}$ are radial  and horizontal 
	component of kinetic energy $E_k$, respectively.  All energies 
	are normalized by $E_k$.
	}
	\label{tab:cnv_nonadi_table}
	\begin{tabular}[htb,width=0.6\columnwidth]{lcccccccr} 
		\hline
		Mode   & $\omega_R$ & $\omega_I$ & $(\alpha\tau_{th}/\tau_{dyn})_{pk}$ & $E_{kr}$ & $E_{kh}$ & $E_p$ & $E_g$ & $E_S$  \\
		\hline
		\hline
		 $$ & $$ & $$ & $$ & $\alpha = 1.0$  \\
		\hline
		 g$^-_0$ & 0.00 & -13.3 & 3.69 & 0.734 & 0.266 & 0.003 & -1.889 & 0.434 \\
		 g$^-_1$ & 0.00 & -9.32 & 3.67 & 0.598 & 0.402 & 0.002 & -1.882 & 0.352 \\
		 g$^-_2$ & 0.00 & -7.36 & 2.90 & 0.521 & 0.479 & 0.001 & -2.140 & 0.442 \\
		 g$^-_3$ & 0.00 & -6.14 & 2.42 & 0.467 & 0.533 & 0.001 & -2.584 & 0.628 \\
		 \hline
		$$ & $$ & $$ &$$ & $\alpha = 0.1$  \\
		\hline
		 g$^-_0$ & 34.0 & -12.0 & 1.42 & 0.629 & 0.371 & 0.029 & -9.041 & 13.82 \\
		 g$^-_1$ & 0.00 & -7.62 & 0.63 & 0.640 & 0.360 & 0.001 & -2.822 & 0.542 \\
		 g$^-_2$ & 0.00 & -6.03 & 0.50 & 0.553 & 0.447 & 0.001 & -3.956 & 0.956 \\
		 g$^-_3$ & 0.00 & -5.02 & 0.60 & 0.494 & 0.506 & 0.001 & -5.887 & 1.667 \\
		  \hline
		$$ & $$ & $$ &$$ & $\alpha = 0.01$  \\
		\hline
		 g$^-_0$ & 48.5 & -7.26 & 0.28 & 0.692 & 0.308 & 0.040 & -11.92 & 18.20 \\
		 g$^-_1$ & 13.1 & -0.33 & 0.16 & 0.508 & 0.492 & 0.004 & -9.118 & 7.147 \\
		 g$^-_2$ & 9.42 & -4.09 & 0.17 & 0.617 & 0.383 & 0.002 & -9.278 & 5.610 \\
		 g$^-_3$ & 0.00 & -3.71 & 0.11 & 0.619 & 0.381 & 0.000 & -17.90 & 5.000 \\
		\hline
	\end{tabular}
\end{table*}
\subsection{Monotonically growing convective mode} 
\begin{figure*}
      \includegraphics[width=1.1\columnwidth,angle=90]{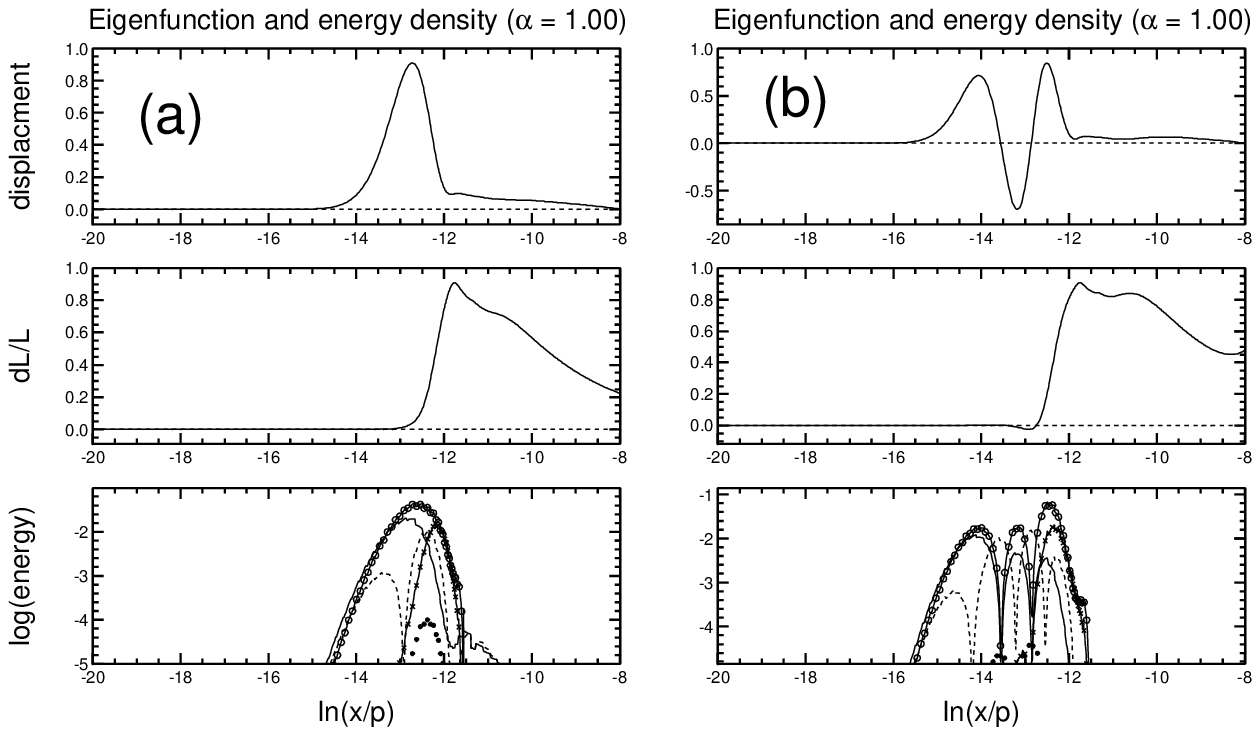} 
    \caption{(a):  Non-adiabatic  displacement, luminosity perturbation, and  energy density of $g^-_0$-mode 
    with $\ell= 6400$ for $\alpha= 1.0$.   
    Solid  and broken lines in the upper panel indicate the real part and the imaginary part of 
    radial displacement, respectively.  Middle panel shows the luminosity perturbation, where ($dL/L$)
    means $dL_{rad}/L_s$. 
    In the lower panel, a line with mark (x)  presents the distribution of entropy energy $e_S$.  The others
    are the same as in Fig.~\ref{fig:cnv_adia_figure}. 
    \newline 
    (b): Non-adiabatic  displacement, luminosity perturbation, and  energy density distribution for $g^-_2$-mode. \\
    {Alt text: Two line graphs. These graphs show the behaviors of monotonically growing convective
    modes. } 
     }
    \label{fig:cnv_nadi_c1_figure}
\end{figure*}

 All modes with $\omega_R=0.00$  grow monotonically as shown in Table~\ref{tab:cnv_nonadi_table},
  although growth rates are reduced comparing with adiabatic case owing to the appearance of
 entropy perturbation ($\delta S$) in  $e_g$ (i.e. Eq.~(\ref{eq:energy_gravity})).  The behaviors of $g^-_0$-
 and $g^-_2$-mode with $\alpha=1.00$ are presented in 
 Fig.~\ref{fig:cnv_nadi_c1_figure}(a) and (b), respectively.  Here the $x$-axis is enlarged to ln(x/p)= [-20.0, -8.0] to
 see the detailed behaviors.  $dL/L$ indicates $\delta L_{rad,r}/L_s$, 
 where $\delta L_{rad,r}$ and $L_s$ are the perturbation of the radiative luminosity 
 at radius $r$ and the total luminosity of the sun, respectively.   Concerning energies, their square terms
 are given in terms of the square of amplitude of complex value. Usually (quasi-adiabatic case) 
 each energy term is calculated as its average  over the period. But the average values have
 no meaning in case of $|\omega_I | \sim \omega_R$. 
 
  For both modes, the radial displacements and the distributions of $e_{kr}$, $e_{kh}$, 
 and $e_{g}$ look like those of adiabatic cases as drawn in Fig.~\ref{fig:cnv_adia_figure}.   
 In fact, a mode can be still classified by number of nodes, which  is also confirmed by 
 number of bumps in the distributions of $e_{kr}$, $e_{kh}$, and $e_g$.   Besides, 
 the imaginary parts of  eigenfunctions  are nearly zero, which means
 quasi-adiabatic condition.  Higher order modes also have smaller growth rates.  These features are common
  in modes with $\omega_R=0.00$.

  Most striking feature is an appearance of  entropy energy distribution ($e_S$) shown in Fig.~\ref{fig:cnv_nadi_c1_figure}.
   A peak of the distribution  of $e_S$ (solid line with x ) corresponds to the steep rise of the perturbation of radiation luminosity $dL/L$  around ln(x/p)= -12.25, at which the divergence of radiation luminosity $dL/L$
(i.e., thermal conduction) is maximum.  In addition to this, the entropy energy $e_S$ has only one 
conspicuous bump.
This is the characteristics of $e_S$ distribution in the non-adiabatic case.
We now define the value of the ratio $\alpha(\tau_{th}/\tau_{dyn})$ at a peak of $e_S$ distribution
denoted by  $(\alpha\tau_{th}/\tau_{dyn})_{pk}$ as a non-adiabaticity measure of  $e_S$ distribution.
 These values are given in Table~\ref{tab:cnv_nonadi_table}.
 
 For monotonically growing convective modes, it should be noted  that a peak of $e_S$  distribution is 
 out of a main bulk of radial
 displacement, and thus the entropy energy  $e_S$ 
  is distributed  only in the partial area of the distributions of $e_k=e_{kr} + e_{kh}$ and $e_g$.  
 In fact, $E_S$ is smaller than $E_g$ by a factor of 4 or 5, and also smaller than $E_k=E_{kr}+E_{kh}$ 
 by a factor  of 2 or 3, as shown in Table~\ref{tab:cnv_nonadi_table}. 
 This fact indicates that the first term ($\xi_r$) in the square bracket of $e_g$ dominates
  over the second term (entropy perturbation) as mentioned in the subsection~\ref{sec:wave_energy}.    
 So the entropy energy $e_S$  does not affect greatly the change of convective mode property, although growth
 rate is depressed.

\subsection{Oscillatory convection} 
 Table~\ref{tab:cnv_nonadi_table} shows that $g^-_0$-mode for $\alpha = 0.1$ and $g^-_0$- , $g^-_1$-, and
  $g^-_2$-mode for $\alpha=0.01$  become oscillatory (i.e., $\omega_R \neq 0$).
In Fig.~\ref{fig:cnv_nadi_c-2_figure}(a) and (b), the behaviors of $g^-_0$- and $g^-_2$-mode  for 
$\alpha=0.01$ are presented, respectively.

They are totally different from the case of monotonically growing mode.  They show only one bump in
the radial displacement and the distributions of energies $e_k$, $e_g$, and $e_S$. 
The distribution of entropy energy $e_S$ (solid line with x ) overlaps almost with that of $e_g$
(solid line with $\bigcirc$), and these two energies prevail against the displacement of kinetic energy $e_k$.
It means that $e_g$ is comparable in magnitude with $e_S$, although the two are different by a
 factor $\nabla/\nabla_{ad}$. 
 In fact,  Table~\ref{tab:cnv_nonadi_table}
shows that $E_S$ is almost
comparable with $E_g$ for oscillatory convective modes except for their signs and  also they are 
considerably larger than $E_{kr}$.   Therefore,  gravity energy $e_g$
acts as a source term of oscillatory convection and entropy energy $e_S$ behaves like its potential energy,
 as stated  in the subsection~\ref{sec:wave_energy},

Oscillatory convection  is no longer  classified by the number of nodes, as the imaginary part of an 
eigenfunction is comparable with its real part.  Instead, the real part of an eigenfrequency 
($\omega_{R}$) may be used as classification, as discussed later in the subsection~\ref{sec:locus}. 

Now what mechanism makes a monotonically growing convective mode oscillatory is raised.
 The most likely mechanism is considered to be
Cowling mechanism \citep{cowling1957}. According to his physical picture, an element displaced upwards
has higher temperature than the surrounding due to super-adiabaticity ($\nabla > \nabla_{ad}$),
   The thermal loss through radiation in this phase forces the element 
to be cooler and denser
in the following descending phase.  The buoyancy accelerates the descending motion
and brings the element deeper until the buoyant force pushes back an element 
owing to thermal gain through radiation from the surrounding.   Although actual force is buoyant force, 
thermal conduction through radiation play a role of preventing an element from monotonically going  
upwards or downwards.    Cowling's explanation is not inconsistent with our
calculations and so it is considered to be the most promising mechanism at the moment. 

\begin{figure*}
	\includegraphics[width=0.9\columnwidth,angle=90]{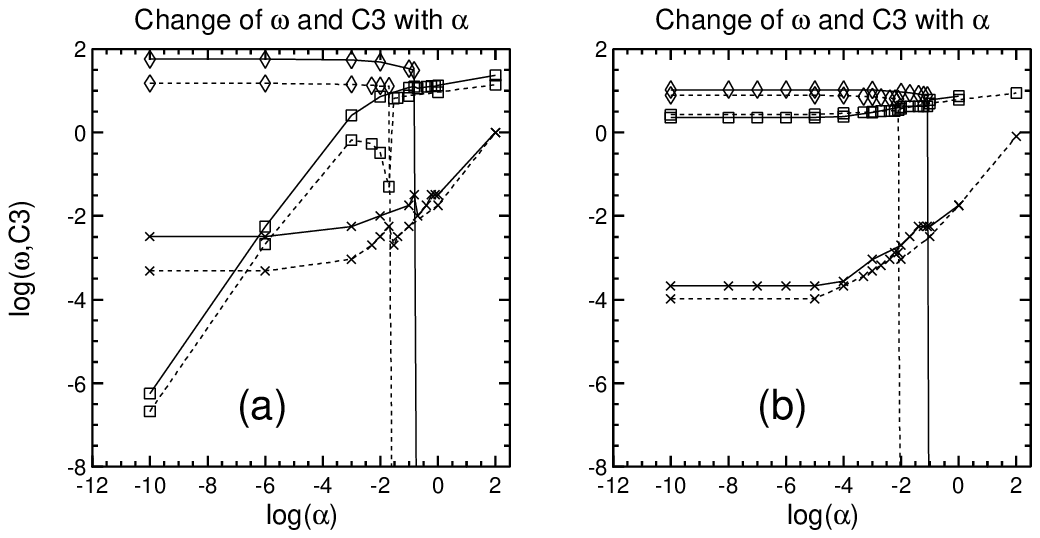} 
    \caption{(a): The changes of $\omega_R$  are indicated by a solid line and a broken line 
        with ($\diamondsuit$)  for  $g^-_0$- and $g^-_1$-mode, respectively. The changes of $\omega_I$  are 
        also drawn by a solid line and a broken line with ($\square$). The lines with (x) show  $C_3$ value 
          (i.e.: $L_{rad}/L_s$) at a peak  of $e_S$ distribution for both modes. 
          \newline 
        (b): The same explanation for $g^-_2$- and $g^-_3$-mode.   \\
        {Alt text: Two line graphs. These graphs describe change of eigenfrequecies with the degree of
        non-adiabaticity. }
        } 
    \label{fig:c2eff_figure}
\end{figure*}
\subsection{Locus of $\omega$ with  $\alpha$}
\label{sec:locus}
We  investigate how  modes ($g^-_0$, $g^-_1$, $g^-_2$, and $g^-_3$) change their
features together with change of $\alpha$. In Fig.~\ref{fig:c2eff_figure}, the  loci of
eigenvalue ($\omega$)  are presented against $\alpha$ for these modes along  with
$C_3$ at a peak of entropy energy distribution ($e_S$),  
where $C_3$ is a fraction of radiation luminosity $L_{rad}$ to
 the total luminosity $L_s$  in the equilibrium model.
 
  As for $\omega_R$ ($\diamondsuit$),  it  becomes  abruptly non-zero ($\omega_R \ne 0$) at a certain $\alpha$
 during decrease of 
  $\alpha$ from  $10^{2}$ to $10^{-10}$, and thereafter keeps almost the same value. Once the thermal 
  conduction through radiation makes a convective mode oscillatory, it does not affect much the frequency itself.
   It  is a very interesting feature, since oscillatory convection 
  can be classified by its frequency. 
  
  Concerning $\omega_I$ ($\square$),  it changes in the two ways. In Fig.~\ref{fig:c2eff_figure}(a) 
  ($g^-_0$ and $g^-_1$), $\omega_I$ is proportional to $\alpha$ in the range  of  $\alpha < 10^{-3}$. 
  On the other hand, it runs    
   horizontally with $\alpha$ as shown in Fig.~\ref{fig:c2eff_figure}(b) ($g^-_2$ and $g^-_3$), 
   although it drops a little bit from 
   the adiabatic value.   Now we pay an attention to the behavior of $C_3$, which is the another coefficient $C_3$ 
   in the energy equation (11) given in \citet{ando1975} other than $C_2$.  ~$C_3$ is a fraction of radiation
   luminosity ($L_{rad}$) to the total luminosity ($L_s$).   The value of $C_3$ at a peak of entropy energy
   distribution $e_S$
    is plotted for each mode by solid lines with (x)  in both figures. ~$C_3$ 
   values in Fig.~\ref{fig:c2eff_figure}(b)
   are less than those in Fig.~\ref{fig:c2eff_figure}(a).  We wonder  that this fact might explain
    the different behavior in $\omega_I$.
   That is, when $C_3$ value is less than a certain value, the thermal conduction through radiation 
   contributing to both of $e_S$ and $e_g$ may be quenched to keep $\omega_I$. 
    On the other hand, for the moderate $C_3$ values, 
    it may be effective to reduce $\omega_I$. Further examination should be stopped here, 
    because a simple numerical experiment with use of $C_3$  is inappropriate owing to its close relation
     with $N^2$-curve through the mixing length theory. 
 
%
\subsection{Application to the present Sun} 
As shown in Table~\ref{tab:cnv_nonadi_table}, an oscillatory convection comes out when the parameter
$(\alpha\tau_{th}/\tau_{dyn})_{pk}$ is less than about 1.0.  Then,
we are aware of a possibility that a monotonically growing convective mode 
becomes oscillatory even in the present Sun. For smaller  $\ell$, $\omega_I$ of a convective mode
decreases  as shown in Fig.~\ref{fig:frequency_figure}, and the above parameter approaches
about one.
 
\begin{figure}
       \includegraphics[width=\columnwidth]{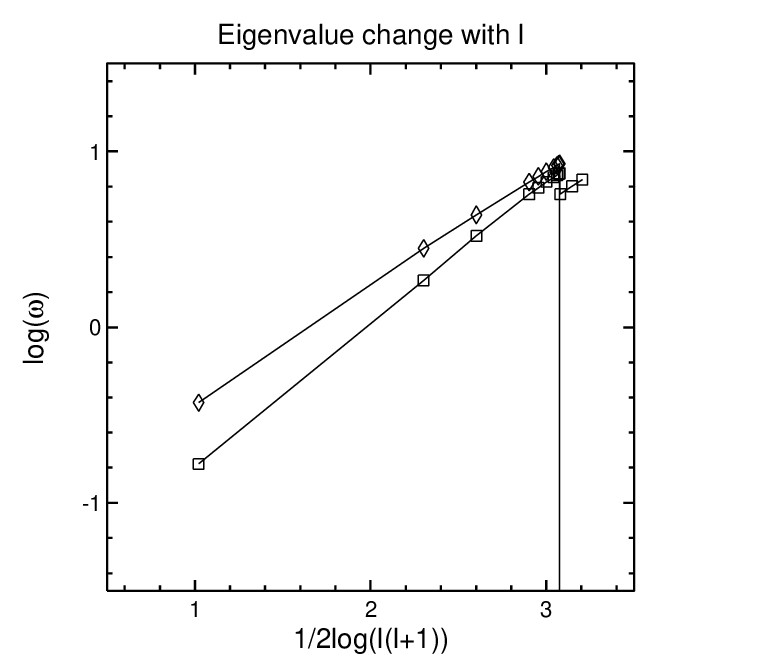} 
    \caption{Eigenvalue  change with $\ell$ for  $g^-_0$-mode.  It becomes
    oscillatory around $\ell$=1200.  \\
    {Alt text: One line graph. This graph shows eigenfrequencies with spherical harmonics index el.}
        } 
    \label{fig:cnv_L10-1600_figure}
\end{figure}

For example,
the behavior of eigenvalue with change of $\ell$ is summarized for $g^-_0$-mode in Fig.~\ref{fig:cnv_L10-1600_figure}.
As $\ell$ decreases from 1600, oscillatory convection appears abruptly around $\ell=1200$,
and then  both of $\omega_R$ and $\omega_I$ decrease.  In the oscillatory phase, 
the fact that  the distribution
of $e_g$ overlaps almost with that of $e_S$  is unchangeable.  

But it should be noted that when  $\omega_R$ for $g^-_0$-mode is less than a peak of the inner $G$-region
(about 6), it might become a mixed mode with a conventional $g$-mode.  In this case, the carful analysis should be required.
But it is beyond the present study,  and we stop here.

\section{Summary and discussion}

In the adiabatic case, the "propagation diagram" for convective modes is introduced based on the local
dispersion relation with $\omega_I$, which is useful to see the trapped region of convective modes.
Any eigenfunction of convective mode is well confined in C-region, which is very similar to
the usual wave like $p$-mode or $g$-mode.   
It is also confirmed that eigenvalue $\omega_I$ is proportional to $\sqrt{\ell(\ell+1)}$, as pointed 
out by \citet{hart1973}.

As for non-adiabatic case, the systematic analysis of convective modes with $\ell=6400$ is carried out
 for three non-adiabatic indicators (i.e., $\alpha= 1.0, ~0.1$, and $0.01$). For $\alpha=1.0$, 
 the convective modes are not so different in eigenfunction ($\xi_r$) and
 energy distributions ($e_k$, $e_g$)  from those of the adiabatic convective modes.   
 The entropy energy $e_S$ plays only a role  of the reduction of growth rate.
 
 For smaller $\alpha$, a monotonically growing convective mode becomes abruptly oscillatory convection 
 when the ratio  $(\alpha\tau_{th}/\tau_{dyn})_{pk}$ of local thermal time scale to dynamical time scale 
at a peak of entropy energy distribution $e_S$ is less than about 1.0.  
Concerning an  oscillatory convection,  the situation is totally different from that of a monotonically growing 
convective mode.   It shows only one bump in the radial displacement $\xi_r$ and the energy distributions of 
energies $e_k$, $e_g$, and $e_S$. 
 The distribution of entropy energy $e_S$ 
almost overlap with the distribution of gravity  energy $e_g$. At the same time, the thermal conduction 
through radiation is almost maximum around a peak of the distribution of entropy energy $e_S$. 
The entropy energy $e_S$ seems to act as the potential energy for oscillatory convection.
The mechanism of making a convective mode oscillatory  may be explained by the Cowling
mechanism proposed by \citet{cowling1957}.   
 Extending this notion, we also show the existence of
oscillatory convection for $\ell<1200$  in the  present Sun.  

Through our systematic analysis, we can understand the role of the gravity energy $e_g$ and
the entropy energy $e_S$ in the following.
When $N^2 < 0$, the entropy energy $e_S$  is considered to play a role of the
potential energy for oscillatory convection, and the gravity energy $e_g$ is source.
On the other hand, in case of $N^2 > 0$
  the gravity energy $e_g$ is the potential energy for $g$-mode, and the entropy energy $e_s$ is source.  
 Therefore, the problem of the overstable convection under the stabilized medium 
  discussed by \citet{cowling1957} and \citet{kato1966} might correspond to
  the latter case.

In this work, the interaction of linear convective mode with  convective motion in the equilibrium state
 has been neglected.  It is not clear at the moment whether such an interaction
 has crucial impact on our result or not.  We  anticipate that
  such a difficult problem will be solved in the future.

\section*{Acknowledgements}

I would like to thank the referee for improving the paper with his helpful comments. 
I am also grateful to National Astronomical Observatory of Japan (NAOJ) for their
financial support.



\bibliographystyle{mnras}    
\bibliography{ando} 

\end{document}